\documentclass[useAMS,usenatbib]{mn2e}
\usepackage{epsfig}

\def\ej{{\rm ej}}

\def\p{{\rm peak}}
\def\iso{{\rm iso}}
\def\jet{{\rm jet}}
\def\sn{{\rm SN}}

\def\sw{{\it Swift}}

\def\dof{{\rm dof}}
\def\nk{{$^{56}{\rm Ni}$}}

\title[Correlation between GRBs and SNe] 
{Correlation between the Peak Spectral Energy of Gamma-Ray Bursts and the
Peak Luminosity of the Underlying Supernovae: Implication for the Nature
of GRB-SN Connection}
\author[Li-Xin Li]{Li-Xin Li\thanks{E-mail: lxl@mpa-garching.mpg.de}\\
Max-Planck-Institut f\"ur Astrophysik, 85741 Garching, Germany}

\begin{document}

\date{}

\date{Accepted 2006 August 15. Received 2006 August 1; in original form 2006 May 1}

\pagerange{\pageref{firstpage}--\pageref{lastpage}} \pubyear{2006}

\maketitle

\label{firstpage}

\begin{abstract}
In this paper we present a correlation between the peak spectral energy 
of gamma-ray bursts (GRBs) and the peak bolometric luminosity of the
underlying supernovae (SNe), based on a sample of four pairs of GRBs-SNe 
with spectroscopically confirmed connection. Combining it with the 
well-known relation between the peak spectral energy and the isotropic 
equivalent energy of GRBs, we obtain an upper limit on the isotropic
energy of GRBs, which is $\approx 10^{52} {\rm erg} \left(L_{\sn,\p}/10^{43}
{\rm erg\, s}^{-1} \right)^{10}$, where $L_{\sn,\p}$ is the peak bolometric 
luminosity 
of the SNe. Our results suggest that the critical parameter determining the 
GRB-SN connection is the peak luminosity of SNe, rather than the feature of 
the SN spectra and/or the SN explosion energy as commonly hypothesized. 
Since it is generally believed that the peak luminosity of SNe powered by
radioactive decays is related to the amount of \nk\, produced in the SN
explosion, the mass of \nk\, may be a key physical factor for understanding 
the nature of GRBs and their connection with SNe. Application of our relation 
to Type Ibc SNe with normal peak luminosities indicates that if those normal
SNe have GRBs accompanying them, the GRBs would be extremely soft and 
sub-energetic in gamma-rays, and hence easier to detect with X-ray or UV 
detectors than with gamma-ray detectors. 
\end{abstract}


\section{Introduction}
\label{intro}

The discovery of SN 1998bw within the error box of GRB 980425 \citep{gal98}
inspired a lot of consideration on the connection between gamma-ray bursts 
(GRBs) and supernovae (SNe). Since then, three more pairs of
GRBs and SNe with spectroscopically confirmed connection have been found,
which are GRB 030329/SN 2003dh \citep{sta03,hjo03}, GRB 031203/SN 2003lw
\citep{mal04,saz04}, and the most recent one discovered by \sw, 
GRB 060218/SN 2006aj \citep{mas06,mod06,cam06,sol06,pia06,mir06,cob06}. 
Interestingly, all of the four SNe are among a special class of Type Ic, 
called the broad-lined SNe, which are characterized by smooth and featureless 
spectra indicating a very large expansion velocity 
\citep[and references therein]{del06,woo06}. Modeling of the SN lightcurves
reveals that the SNe with GRB-connection have very large explosion energy
and mass production of \nk\, compared to normal Type Ibc SNe
\citep{iwa98,nak01,den05,maz06b}, except SN 2006aj which requires an explosion
energy that is comparable to that of normal SNe Ibc \citep{maz06}. 
These facts have motivated people to invent the term ``hypernovae'' for
this special and much more powerful class of SNe (Iwamoto et al. 1998; see
also Paczy\'nski 1998a,b).

A less direct way for identifying the GRB-SN connection is observing the 
rebrightening and/or flattening (called ``red bumps'') in the late GRB 
afterglows, which can be interpreted as the emergence of the underlying SN 
lightcurves \citep[and references therein]{blo99,zeh04,sod05,bers06}. 
Although alternative explanations with dust echoes \citep{esi00} and dust 
sublimation \citep{wax00} have been proposed, several groups have 
successfully fitted SN 1998bw templates to explain the late-time bumps 
\citep{blo02,gar03,gre03,sta05,bers06}. A systematic study on the GRB 
afterglows with this approach by \citet{zeh04} suggests that all long-duration 
GRBs are associated with SNe. 

Despite the exciting developments in the past eight years in the detection 
and observation of GRB-SN connection, by now no any quantitative relation 
between the parameters of GRBs and that of SNe has been found (see, e.g.,
Zeh, Klose \& Hartmann 2004; Ferrero et al. 2006), although it is commonly 
conceived that only very bright SNe can produce GRBs, based on the fact that 
all SNe with confirmed GRB-connection are much brighter than average and that 
the rate of GRBs and ``hypernovae'' are several orders of magnitude lower 
than the rate of core-collapse SNe \citep{pod04}. The lack of a quantitative 
relation between GRBs and SNe has frustrated the advance in understanding 
the nature of GRBs, although many people believe that long-duration GRBs are 
produced by the core-collapse of massive stars \citep{mac01,woo06a}.

In this paper, we present the discovery of a quantitative relation between
the GRBs and the underlying SNe, based on the observational data of the
four pairs of GRBs-SNe with spectroscopically confirmed connection 
(Table~\ref{grbsn}). We show that, the peak spectral energy of the GRB is 
strongly correlated with the peak bolometric luminosity of the SN. Then, 
combining with the relation between the peak spectral energy and the 
isotropic equivalent energy of GRBs found by \citet{ama02}, we explore the 
implications of the correlation that we have found for the GRB-SN connection 
and for the nature of GRBs.

\begin{table*}
\centering
\begin{minipage}{160mm}
\caption{Gamma-ray bursts and supernovae with spectroscopically confirmed connection$^\dagger$}
\label{grbsn}
\begin{tabular}{llllllll}
\hline
GRB/SN\hspace{1cm} & $z^{\rm (a)}$\hspace{0.48cm} & $E_{\gamma,\p}^{\rm
(b)}$\hspace{0.33cm} & $E_{\gamma,\iso}^{\rm (c)}$\hspace{1.4cm} &
$M_{\sn,\p}^{\rm (d)}$\hspace{0.76cm} & $E_K^{\rm (e)}$\hspace{0.84cm} &
$M_\ej^{\rm (f)}$\hspace{0.46cm}  & $M_{\rm Nickel}^{\rm (g)}$ \\
\hline
980425/1998bw & $0.0085$ & $55\pm 21$ & $0.00009\pm 0.00002$ & $-18.65\pm 0.20$
& $5.0\pm 0.5$ & $10\pm 1$ & $0.38$--$0.48$ \\
030329/2003dh & $0.1687$ & $79\pm 3$  & $1.7\pm 0.2$ & $-18.79\pm 0.23$ &
$4.0\pm 1.0$ & $8\pm 2$ & $0.25$--$0.45$ \\
031203/2003lw & $0.1055$ & $159\pm 51$ & $0.009\pm 0.004$ & $-18.92\pm 0.20$ &
$6.0\pm 1.0$ & $13\pm 2$ & $0.45$--$0.65$ \\
060218/2006aj & $0.0335$ & $4.9\pm 0.4$ & $0.0059\pm 0.0003$ & $-18.16\pm 0.20$
& $0.2\pm 0.02$ & $2\pm 0.2$ & $0.2\pm 0.04$ \\
\hline
\end{tabular}
\begin{list}{}{}
\item[$^\dagger$] Following \citet{maz06b}, we assume $H_0 = 72$ km s$^{-1}$ 
Mpc$^{-1}$, $\Omega_m = 0.28$, and $\Omega_\Lambda = 0.72$. 
Quantities calculated in a different cosmology are converted to this 
cosmology.
\item[$^{\rm (a)}$] Cosmic redshift. From \citet{gal98} (GRB 980425/SN 1998bw),
\citet{sta03} (GRB 030329/SN 2003dh), \citet{pro04} (GRB 031203/SN 2003lw),
\citet{pia06} and \citet{mir06} (GRB 060218/SN 2006aj).
\item[$^{\rm (b)}$] Peak energy of the integrated GRB spectrum in units of 
keV, measured in the GRB frame. From \citet{yam03} (GRB 980425), \citet{sak05}
(GRB 030329), \citet{ula05} (GRB 031203), and \citet{cam06} (GRB 060218). 
\item[$^{\rm (c)}$] Isotropic equivalent energy of the GRB in units of 
$10^{52}$ erg, defined in the $1-10,000$ keV energy band in the GRB frame. 
From \citet{cam06} (GRB 060218), and \citet{ama06} (other bursts).
\item[$^{\rm (d)}$] Peak bolometric magnitude of the supernova, defined 
in the $3,000-24,000$ \AA\, wavelength band in the SN frame. From 
\citet{maz06b} (SN 1998bw, SN 2003lw), \citet{pia06} (SN 2006aj), and 
\citet{den05} (SN 2003dh, and Appendix~\ref{sn03dh} of this paper).
\item[$^{\rm (e)}$] Explosion kinetic energy of the supernova in units of 
$10^{52}$ erg. From \citet{maz06} (SN 2006aj, error of 10 percent added), 
and \citet{maz06b} (other SNe).
\item[$^{\rm (f)}$] Ejected mass in the supernova explosion in units of 
$M_\odot$. From \citet{maz06} (SN 2006aj, error of 10 percent added), and 
\citet{maz06b} (other SNe).
\item[$^{\rm (g)}$] Mass of \nk\, produced by the SN explosion in units 
of $M_\odot$. From \citet{maz06} (SN 2006aj, the error corresponds to a 
0.2 mag error in magnitude), and \citet{maz06b} (other SNe).
\end{list}
\end{minipage}
\end{table*}

\section{The Peak Spectral Energy of GRBs versus the Peak Bolometric 
Luminosity of Supernovae}
\label{peaks}

Informations of the four pairs of GRBs and SNe with spectroscopically 
confirmed connection are summarized in Table~\ref{grbsn}, including their 
cosmic redshift, the peak spectral energy and isotropic equivalent energy 
of the GRBs, the peak bolometric magnitude, explosion energy, ejected mass, 
and nickel yield of the SNe. Following \citet{maz06b}, we assume a cosmology 
with $H_0 = 72$ km s$^{-1}$ Mpc$^{-1}$, $\Omega_m = 0.28$, and 
$\Omega_\Lambda = 0.72$. All quantities calculated in a different cosmology 
are converted to the above cosmology.

Among the four bursts, GRB 030329 is the brightest one in terms of the 
isotropic equivalent energy (or the peak luminosity). However, the supernova 
associated with it, SN 2003dh, is not most powerful. Its total explosion 
kinetic energy is exceeded by that of SNe 1998bw and 2003lw, associated with 
GRBs 980425 and 031203 
respectively. In terms of the bolometric luminosity, SN 2003dh is also fainter 
than SN 2003lw. Although GRB 030329 is very bright and energetic compared to 
the other three GRBs, it is significantly weaker than average long-duration 
GRBs.

GRB 980425, the nearest burst with measured redshift to date and the first 
GRB that has been discovered to be connected to a SN, is least energetic in 
terms of the isotropic equivalent energy. However, the supernova associated 
with it, SN 1998bw, is very powerful and very bright. GRB 031203, associated 
with SN 2003lw, is analogous to GRB 980425 in many aspects \citep{sod04}. 
It is also underluminous and has a very bright and powerful supernova. 
However, as can be seen from Table~\ref{grbsn}, GRB 031203 has a much harder
spectrum than GRB 980425 as indicated by its much larger peak spectral
energy.

GRB 060218, recently discovered and the second nearest one, is a
very peculiar burst \citep{pia06,cam06}. It has an extremely long duration 
($\approx 2,000$ sec), and an extremely soft spectrum (with a peak spectral 
energy $E_{\gamma,\p} \approx 4.9$ keV in the GRB frame). It is also 
sub-energetic, has an isotropic equivalent energy similar to that of 
GRB 031203. The supernova associated with it, SN 2006aj, is the faintest and 
the least powerful one among the four GRB-connected SNe. The modeling by 
\citet{maz06} reveals that it has an explosion kinetic energy $E_K \approx 
2\times 10^{51}$ erg, ejected mass $M_\ej \approx 2 M_\odot$, and ejected 
\nk\, $\approx 0.2 M_\odot$. Although SN 2006aj is much brighter than average 
SNe Ic and has a much smoother spectrum, its explosion appears to be
less powerful than other GRB-connected SNe but closer to normal SNe Ic.

\begin{figure}
\vspace{2pt}
\includegraphics[angle=0,scale=0.467]{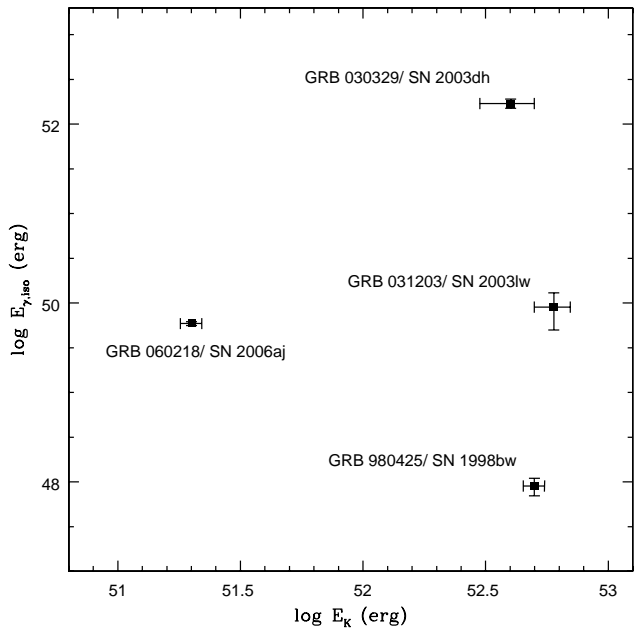}
\caption{The isotropic equivalent energy of GRBs versus the explosion 
kinetic energy of the underlying SNe. Clearly, there is no correlation 
between them.
}
\label{ek_eiso}
\end{figure}

Despite the very narrow distribution in the peak bolometric magnitudes of the 
four SNe, from $-18.16$ to $-18.92$ mag [corresponding to a factor of 2 
variation in the peak bolometric luminosity, $(0.559 - 1.13)\times 10^{43}$ 
erg s$^{-1}$], the distribution in the isotropic energy of the GRBs is 
extremely wide, $(0.0001 - 1.7)\times 10^{52}$ erg. It appears that there 
does not exist a relation between the isotropic equivalent energy of the 
bursts and the explosion energy of the supernovae, as can be seen from 
Fig.~\ref{ek_eiso}.  

\begin{figure}
\vspace{2pt}
\includegraphics[angle=0,scale=0.467]{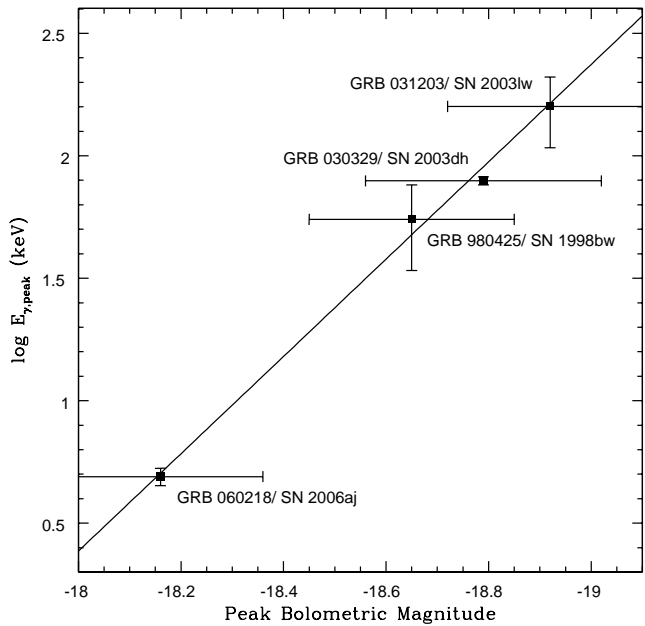}
\caption{The peak spectral energy of GRBs versus the peak 
bolometric magnitude of the underlying SNe. The straight line is a
least-$\chi^2$ fit to the data (eq.~\ref{ep_mp}).
}
\label{mag_epeak}
\end{figure}

However, there appears to be a very good correlation between the peak spectral
energy of the GRB (defined in the GRB frame) and the peak bolometric magnitude
(luminosity) of the SN, as shown in Fig.~\ref{mag_epeak}. Despite the large
systematic errors in the peak bolometric magnitude relative to its narrow  
distribution, a correlation between $M_{\sn,\p}$ (the peak bolometric 
magnitude of the SN) and $E_{\gamma,\p}$ (the peak spectral energy of the GRB) 
is remarkable. The Pearson linear correlation coefficient between $-M_{\sn,
\p}$ and $\log E_{\gamma,\p}$ is calculated to be $r = 0.997$, corresponding 
to a probability $P = 0.003$ for zero correlation. This indicates that 
$-M_{\sn,\p}$ and $E_{\gamma,\p}$ are strongly correlated. (For comparison, 
the Pearson linear correlation coefficient between the log explosion energy 
and the log isotropic energy in Fig.~\ref{ek_eiso} is $r = 0.019$, 
corresponding to a probability $P = 0.981$ for zero correlation.)

A least-$\chi^2$ linear fit to $M_{\sn,\p} - \log E_{\gamma,\p}$, taking into
account the errors in both variables,  gives
\begin{eqnarray}
	\log E_{\gamma,\p} = -35.38 - 1.987\, M_{\sn,\p} 
	\label{ep_mp}
\end{eqnarray}
with $\chi^2/\dof = 0.02$, where $E_{\gamma,\p}$ is in keV. This relation 
is equivalent to
\begin{eqnarray}
	E_{\gamma,\p} = 90.2\,{\rm keV} \left(\frac{L_{\sn,\p}}{10^{43} 
		{\rm erg\,s}^{-1}}\right)^{4.97} \;,
	\label{ep_lp}
\end{eqnarray}
where $L_{\sn,\p}$ is the peak bolometric luminosity of the supernova defined 
in the $3,000-24,000$ \AA\, wavelength band in the SN frame.

\begin{figure}
\vspace{2pt}
\includegraphics[angle=0,scale=0.467]{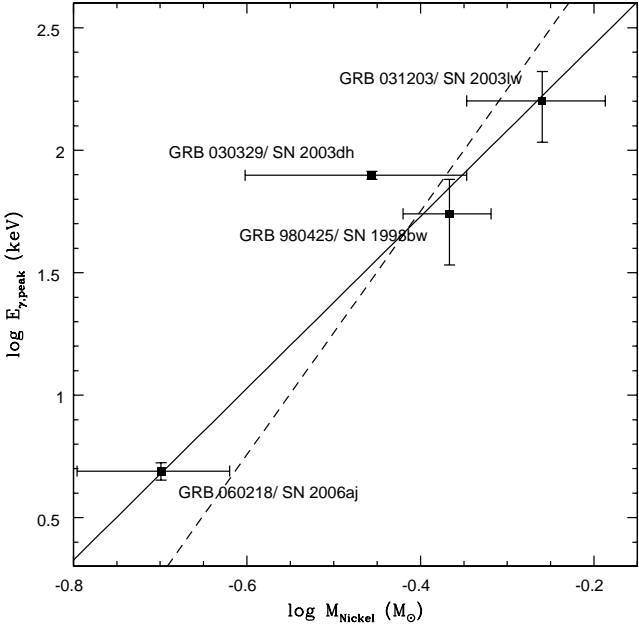}
\caption{The peak spectral energy of GRBs versus the mass of \nk\, generated
in the underlying SNe. For SNe 1998bw, 2003dh, and 2003lw, the value of 
$M_{\rm Nickel}$ is taken to be the mean of the upper and lower limits
in Table~\ref{grbsn}. The solid straight line is a least-$\chi^2$ fit to the 
data,
$\log E_{\gamma,\p} = 3.13 + 3.51 \log M_{\rm Nickel}$ with $\chi^2/\dof = 
0.4$. If the slope is fixed at $4.97$ (i.e., assuming that eq.~\ref{ep_lp}
holds and the nickel yield is proportional to the peak luminosity of the SNe),
a least-$\chi^2$ fit leads to $\log E_{\gamma,\p} = 3.74 + 4.97 
\log M_{\rm Nickel}$ with $\chi^2/\dof = 1.03$ (the dashed line).
}
\label{n56_epeak}
\end{figure}

It is well known that the peak luminosity of SNe powered by radioactive 
decays is related to the mass of \nk\, generated in the SN ejecta 
\citep{arn82,mae03,nom04}. Approximately, the maximum luminosity is 
proportional to the  mass of \nk. But it also depends on the diffusion time 
of the photons generated by the deposition of the gamma-rays emitted by the 
decay of freshly synthesized \nk\ to $^{56}$Co and hence to stable 
$^{56}$Fe \citep{maz06}. To check the relation between the peak spectral 
energy of GRBs and the mass of \nk\, produced by the SNe, in 
Fig.~\ref{n56_epeak} we plot $E_{\gamma,\p}$ against $M_{\rm Nickel}$, the 
mass of \nk. Not surprisingly, $E_{\gamma,\p}$ is also correlated with 
$M_{\rm Nickel}$, although the correlation is not as tight as that in 
$E_{\gamma,\p}-M_{\sn,\p}$ in Fig.~\ref{mag_epeak}. The Pearson linear 
correlation coefficient between $\log E_{\gamma,\p}$ and 
$\log M_{\rm Nickel}$ is $r = 0.95$, corresponding to a probability 
$P=0.05$ for zero correlation.

Although the mass of \nk\, is a parameter that is more physical than the 
peak luminosity, in this paper we focus on the relation between the peak
spectral energy of GRBs and the peak luminosity of SNe since the peak 
luminosity is a directly measurable quantity. Unlike the mass of \nk, the 
peak luminosity does not depend on the SN model and hence does not suffer the 
errors from the model assumptions.

\section{The Energetic Nature of GRBs associated with Supernovae}
\label{sub_energetic}

It has been found that the isotropic equivalent energy of long-duration GRBs, 
defined in the $1-10,000$ keV band in the GRB frame, is correlated with the 
peak energy of the integrated spectra of GRBs, with only a few outliers 
\citep{ama02,ama06}. The correlation is even better when the correction to 
the GRB energy from jet collimation is included \citep{ghi04}. A recent 
study with an updated GRB sample consisting of 41 long GRBs by \citet{ama06} 
gives, when normalized to the cosmology adopted in this paper and outliers 
are excluded,
\begin{eqnarray}
	E_{\gamma,\p} = 97\,{\rm keV} \left(\frac{E_{\gamma,\iso}}{10^{52} 
		{\rm erg}}\right)^{0.49} \;.
	\label{ep_ei}
\end{eqnarray}

\begin{figure}
\vspace{2pt}
\includegraphics[angle=0,scale=0.46]{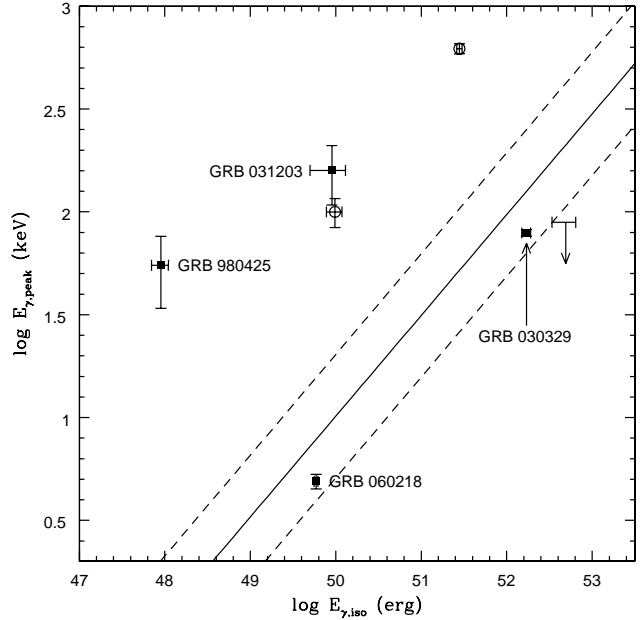}
\caption{The peak spectral energy versus the isotropic equivalent energy, 
for the four GRBs with SN-connection. The solid line is the relation in
eq.~(\ref{ep_ei}), the best power-law fit to 41 long-duration GRBs 
\citep{ama06}. The two dashed lines delineate the region of a logarithmic 
deviation of $0.3$ (2-$\sigma$) in $E_{\gamma,\p}$. The two open circles 
are short-duration GRBs 050709 and 051221a. The downward arrow on the
right shows the upper bound on the $E_{\gamma,\p}$ of GRB 050315 (see the 
text).
}
\label{eiso_epeak}
\end{figure}

GRBs 030329 and 060218 are consistent with the relation in
equation~(\ref{ep_ei}), but 980425 and 031203 are not, see
Fig.~\ref{eiso_epeak}. Among well-studied long GRBs, 980425 and 031203
are indeed the only known outliers to the $E_{\gamma,\p} - E_{\gamma,\iso}$ 
relation \citep{ama06b}.

It appears that all GRBs that violate the $E_{\gamma,\p} - E_{\gamma,\iso}$ 
relation stay on the side of having smaller isotropic energy than predicted 
by the relation, see Fig.~\ref{eiso_epeak}. However, there is one possible 
exception: GRB 050315 at redshift $1.949$, a bright long burst discovered 
by \sw\,. \citet{vau06} estimated that for this burst the peak spectral energy
is $\la 30$ keV in the observer frame, i.e., $E_{\gamma,\p} \la 89$ keV in 
the GRB frame. This low value of $E_{\gamma,\p}$ makes GRB 050315 marginally 
violate the $E_{\gamma,\p} - E_{\gamma,\iso}$ relation by having a slightly 
larger isotropic energy (Fig.~\ref{eiso_epeak}). However, in obtaining their
result, \citet{vau06} have assumed a too large absolute value for the photon 
index of low energy, $\alpha = -1.88$. If taking $\alpha = -1.3$, they 
obtained a larger upper bound for the peak spectral energy ($\la 43$ keV 
in the observer frame), making GRB 050315 closer to the $E_{\gamma,\p} - 
E_{\gamma,\iso}$ relation. The most likely value of $-\alpha$ for 
GRBs observed by BATSE/{\it CGRO} was $1$ \citep{pre00}, much smaller than 
the value that was assumed by Vaughan et al. Hence, because of the fact that 
the low energy photon index of GRB 050315 cannot be determined with the 
BAT/\sw\, data alone, we would not consider GRB 050315 as a serious case that 
violates the $E_{\gamma,\p} - E_{\gamma,\iso}$ relation. 

Then, based on the data of GRBs that have accurately determined peak spectral 
energy and isotropic equivalent energy, we can say with fair confidence that 
the $E_{\gamma,\p} - E_{\gamma,\iso}$ relation, i.e., equation~(\ref{ep_ei}), 
gives a fairly accurate estimate on the isotropic energy for normal GRBs, and 
an upper bound on the isotropic energy for sub-energetic 
GRBs.\footnote{\citet{nak05} showed that at least 25\% of the BATSE GRBs are
outliers to the $E_{\gamma,\p} - E_{\gamma,\iso}$ relation, and suggested that
eq.~(\ref{ep_ei}) should be considered as an upper bound on the isotropic
energy of GRBs.} Then, the combination of equations~(\ref{ep_lp}) and 
(\ref{ep_ei}) leads to
\begin{eqnarray}
	E_{\gamma,\iso} \la 0.86 \times 10^{52} {\rm erg} \left(
		\frac{L_{\sn,\p}}{10^{43} {\rm erg\,s}^{-1}}
		\right)^{10} \;.
	\label{eiso_up}
\end{eqnarray}

Equation~(\ref{eiso_up}) provides a strong constraint on the isotropic
equivalent energy of GRBs associated with SNe. Because of the very steep 
slope in $\log E_{\gamma,\iso} - \log L_{\sn,\p}$, equation~(\ref{eiso_up}) 
describes the fact that the isotropic energy of GRBs is distributed in an 
extremely wide range while the peak luminosity of the underlying SNe has an 
extremely narrow distribution.

\section{The Mildly-Relativistic Nature of GRBs with Soft Spectra}
\label{sub} 

A common feature of the four SN-connected GRBs is that all of them are soft,
characterized by their small peak spectral energy compared to normal 
cosmological GRBs. An analysis by \citet{ama06} on 45 GRBs with 
well-determined peak spectral energy shows that $E_{\gamma,\p}$ can be 
described by a log-normal distribution with a mean $\sim 350$ keV and a 
logarithmic dispersion of $\sim 0.45$. The hardest one in the four 
SN-connected GRBs, 031203, has a peak spectral energy $\approx 159$ keV, 
smaller than the mean of the distribution but still within 1-$\sigma$. While 
the softest one, GRB 060218, has a peak spectral energy as small as $\approx 
4.9$ keV, deviating from the mean by $4$-$\sigma$.

\begin{figure}
\vspace{2pt}
\includegraphics[angle=0,scale=0.467]{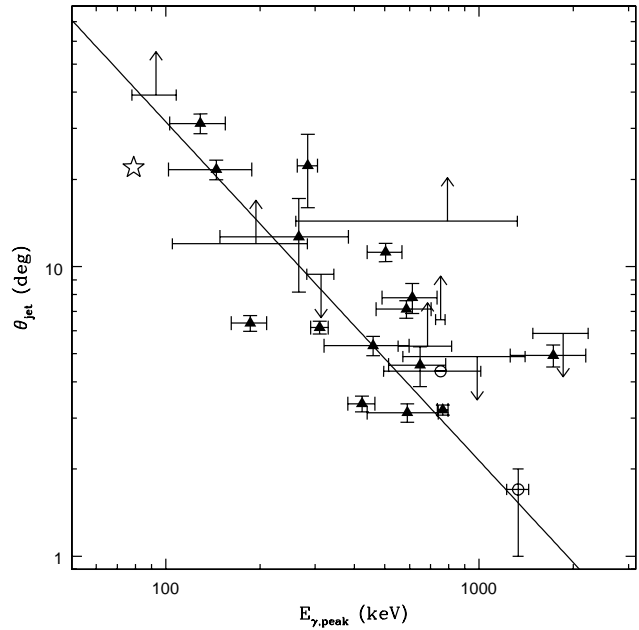}
\caption{The jet opening angle of GRBs versus the peak energy of their 
spectra measured in the GRB frame. Filled triangles are 15 GRBs that 
have accurately determined jet opening angles, taken from \citet{blo03} 
(excluding GRB 000301c whose peak spectral energy is not available). 
Upward (downward) arrows are lower (upper) limits on the opening angles 
for 8 bursts from the same paper. Two open circles are GRB 050603 
\citep{ber05,gru06} and GRB 051022 (without an error bar in $\theta_\jet$) 
\citep{rac05}. The peak spectral energy of all GRBs was taken from 
\citet{ama06}. Whenever they are available, error bars are indicated. The 
straight line is a maximum-likelihood fit to the data (eq.~\ref{theta_ep}), 
excluding the 8 GRBs with only limits. [The star is GRB 030329 (not included 
in the fit), whose jet opening angle is $\sim 22$ degree as inferred from 
radio observation \citep{hor05}].
}
\label{ep_theta}
\end{figure}

The peak spectral energy of GRBs is anti-correlated with the jet opening
angle \citep{lam05}. In Fig.~\ref{ep_theta} we plot the jet opening angle 
at the time of jet break, versus the peak spectral energy of the burst for 
26 GRBs (see the figure caption for the sources of data). All 
the opening angles were calculated from the time of jet break in the 
afterglows, except that of GRB 030329---the only SN-connected GRB included 
in the plot and marked by a star---which was obtained less directly by 
modeling the radio afterglow. With GRB 030329 and those bursts with only 
limits on opening angles being excluded (then we had 17 GRBs left), we 
obtained a maximum-likelihood fit to the data 
\begin{eqnarray}
	\log \theta_\jet = 3.84 - 1.17\, \log E_{\gamma,\p} \;,
	\label{theta_ep}
\end{eqnarray}
where $E_{\gamma,\p}$ is in keV and $\theta_\jet$ is in degree. This relation
is not sensitive to the assumed cosmology, since the jet opening angle weakly
depends on the luminosity distance \citep{sar99,fra01,blo03}.

Thus, a smaller value of the peak spectral energy indicates a larger jet 
opening angle, and hence a smaller Lorentz factor $\Gamma$ since $\Gamma\sim
\theta_\jet^{-1}$ at the time of jet break. For a GRB with a very small 
peak spectral energy, the Lorentz factor of its outflow must be very 
small compared to typical GRBs whose Lorentz factors have been argued to be 
$\ga 300$ based on the fact of the presence of MeV photons in their spectra 
\citep{pir04}. Hence, GRBs with soft spectra must be mildly-relativistic, 
where by ``mildly-relativistic'' we mean that the Lorentz factor $\Gamma < 
100$.

Given that $E_{\gamma,\p} \approx 159$ keV for GRB 031203, from 
Fig.~\ref{ep_theta} its jet opening angle would be in the range of $10 - 30$ 
degree. While for GRB 980425 with $E_{\gamma,\p}\approx 55$ keV, the 
relation~(\ref{theta_ep}) predicts a jet opening angle $\sim 60$ degree. For 
GRB 060218, who has the smallest peak energy $E_{\gamma,\p}\approx 4.9$ keV, 
the relation~(\ref{theta_ep}) predicts that its jet opening angle would be 
$1,000$ degree! Hence, the anti-correlation between the peak spectral energy 
and the jet opening angle indicates that GRB 060218 is almost perfectly 
spherical.

Equation~(\ref{theta_ep}) suggests that all GRBs (at least those of long
duration) with peak spectral energy $\la 40$ keV (in the GRB frame) are 
spherical and only mildly-relativistic since then the predicted jet opening 
angle $\ga 90$ degree.

A popular explanation for sub-energetic GRBs has been that they are normal 
GRBs viewed away from their jet axes 
\citep[and references therein]{wax04,ram05}, but our results in this section
suggest that sub-energetic GRBs are spherical and hence intrinsically faint. 
Our view is supported by the radio observations on GRBs 980425, 
031203, and 060218 \citep{sod04a,sod04,sod06}. For example, the radio 
afterglow lightcurve of GRB 060218 did not show a signature of jet break 
after 22 days of the burst, indicating that the jet opening angle 
$\theta_\jet > 1.4\,{\rm rad} \approx 80\,{\rm degree}$ \citep{sod06}. 
The fact that the rate of low-luminosity GRBs exceeds that expected from 
off-axis models by at least a factor of ten also suggests that 
low-luminosity GRBs are intrinsically sub-energetic \citep{cob06,sod06,lia06}.

\section{Implication for the Nature of GRB-SN Connection}
\label{implication}

Although it is always a risk to extend a relation beyond the range based on 
which the relation was derived, we cannot resist to apply the relations 
derived in previous sections (eqs.~\ref{ep_lp} and \ref{eiso_up}) to normal 
Type Ibc SNe and to cosmological GRBs and see where the relations lead us to 
and if the results contradict observations.

In the Fig.~3 of \citet{pia06}, the brightest supernova next to SN 2006aj 
is the ``standard'' Type Ic SN 1994I in the spiral galaxy M51 with a distance 
$8.4\pm 0.6$ Mpc from us \citep{fel97}. The peak bolometric luminosity of 
SN 1994I is $\approx 2.34 \times 10^{42} {\rm erg}\,{\rm s}^{-1}$ 
\citep{sau06}, fainter than SN 1998bw by 1.4 mag. By equation~(\ref{eiso_up}), 
if there was a GRB associated with SN 1994I, its isotropic energy would be 
$\la 4 \times 10^{45}$ erg, smaller than that of GRB 980425 by two orders of 
magnitude. Although SN 1994I is four times closer to us than 
SN 1998bw/GRB 980425, the burst related to SN 1994I would still be ten times 
fainter than GRB 980425 in gamma-rays if it had a similar duration. The peak 
spectral energy of the burst inferred from equation~(\ref{ep_lp}) is 
$\approx 0.07$ keV, in the soft X-ray and extreme UV band. 

Applying equations~(\ref{ep_lp}) and (\ref{eiso_up}) to SN 1997ef 
\citep{iwa00,maz00,maz04,pia06} and SN 2002ap \citep{maz02,tom06}, which 
have been 
classified as ``hypernovae'' by the similarity of their spectra to that of 
SN 1998bw and their large explosion energy, we obtain $E_{\gamma,\p}\approx 
0.017$ keV, $E_{\gamma,\iso} \la 2.7\times 10^{44}$ erg for SN 1997ef, and 
$E_{\gamma,\p} \approx 0.016$ keV, $E_{\gamma,\iso} \la 2.3\times 10^{44}$ 
erg for SN 2002ap. The peak spectral energy of the potential bursts is in 
the UV band, and the isotropic gamma-ray energy is smaller than that of 
GRB 980425 by more than three orders of magnitude.

SN 1997ef, occurred in UGC 4107, has a mass of \nk\, that is about twice
that in other SNe with similar brightness because of its very late peak 
\citep{maz00,maz04,iwa00}. Converted to the cosmology adopted in this paper, 
its $M_{\rm Nickel} \approx 0.13 M_\odot$. Then, the $E_{\gamma,\p}-
M_{\rm Nickel}$ relation found in Sec.~\ref{peaks} (the solid straight line 
in Fig.~\ref{n56_epeak}) gives $E_{\gamma,\p}\approx 1$ keV, and hence
$E_{\gamma,\iso} \la 0.9\times 10^{48}$ erg by equation~(\ref{ep_ei}). That
is, the upper limit of the isotropic energy of the burst associated with
SN 1997ef suggested by the $E_{\gamma,\p}-M_{\rm Nickel}$ relation is 
comparable to that of GRB 980425. The $E_{\gamma,\p}-M_{\rm Nickel}$ relation
leads to larger values of $E_{\gamma,\p}$ and $E_{\gamma,\iso}$ than the 
$E_{\gamma,\p}-M_{\sn,\p}$ relation, resulted from the smaller slope in 
$E_{\gamma,\p}-M_{\rm Nickel}$ (Fig.~\ref{n56_epeak}, the solid line versus 
the dashed line).

SN 1997ef has been suggested to be associated with GRB 971115 by the fact 
that the two may be compatible with each other in position and time of 
occurrence \citep{wan98}. However, the correlation is much weaker than 
that in the 
case of SN 1998bw/GRB 980425. SN 1997ef was slightly outside the 2-$\sigma$ 
error box of GRB 971115, and the angular separation between them was as large 
as $25$ degree. The temporal association was also weak: the maximum of the
optical lightcurve of SN 1997ef was delayed from GRB 971115 by about 20 days, 
in contrast to the 9-17 days for the four spectroscopically confirmed
SNe-GRBs. The explosion date of SN 1997ef was estimated to be November $20 
\pm 1$ day \citep{maz00}, delayed from GRB 971115 by $5\pm 1$ day which 
is much longer than typical SN-GRB time lags \citep{del06}.

As SN 1994I, SN 2002ap is another nearby supernova, discovered in M74 with a 
distance $7.8^{+0.4}_{-0.7}$ Mpc. An intensive search of all available 
gamma-ray data obtained between January 21 and January 29 of 2002 for
bursts that could be localized by the Interplanetary Network (IPN) found no 
GRB associated with SN 2002ap (Hurley et al. 2002; see, however, Gal-Yam, 
Ofek \& Shemmer 2002). The peak bolometric luminosity of SN 2002ap is 
$\approx 1.75\times 10^{42}$ erg s$^{-1}$, fainter than SN 1998bw by 1.75 
mag \citep{tom06}. Despite its closer distance, our relation predicts 
that the burst associated with SN 2002ap would look $\sim 190$ times fainter 
than GRB 980425 in gamma-rays.

SN 2004aw is one of the most well observed Type Ic supernovae, discovered in
a tidal tail of a barred spiral galaxy NGC 3997 at redshift $z=0.0163$ 
\citep{tau06}. It is intrinsically slightly brighter than SN 1994I, but 
fainter than SN 1998bw by 1.3 bolometric magnitude at peak. The optical 
spectrum of SN 2004aw bridges a normal SN Ic like SN 1994I and the group of 
broad-lined SNe Ic. No GRB has been found to be associated with SN 2004aw
\citep{tau06}. 
Submitting the peak bolometric luminosity of SN 2004aw ($\approx 2.63\times 
10^{42}$ erg s$^{-1}$, Taubenberger et al. 2006) into equations~(\ref{ep_lp}) 
and (\ref{eiso_up}), we get $E_{\gamma,\p} \approx 0.12$ keV and 
$E_{\gamma,\iso} \la 1.4\times 10^{46}$ erg. Given its distance of $68.2$ 
Mpc, the potential GRB associated with SN 2004aw would look at least 200 
times fainter than GRB 980425.

Therefore, if normal Type Ibc SNe are accompanied by GRBs, the GRBs should 
be extremely under-luminous in the gamma-ray band despite their close
distances. Their peak spectral energy is expected to be in the soft X-ray 
and UV band, so they may be easier to detect with a X-ray or UV detector than
with a gamma-ray detector. We note that in terms of both total energy and 
photon energy, the bursts are similar to the shock breakout flashes 
predicted for Type Ibc SNe \citep{bli02,li06}. 
Flashes from shock breakout in SNe were first predicted by \citet{col68} 
almost forty years ago, originally proposed for GRBs that had not been 
discovered yet. However, they have never been unambiguously detected in 
supernova observations because of their short duration compared to SNe 
\citep{cal04}.

\begin{figure}
\vspace{2pt}
\includegraphics[angle=0,scale=0.467]{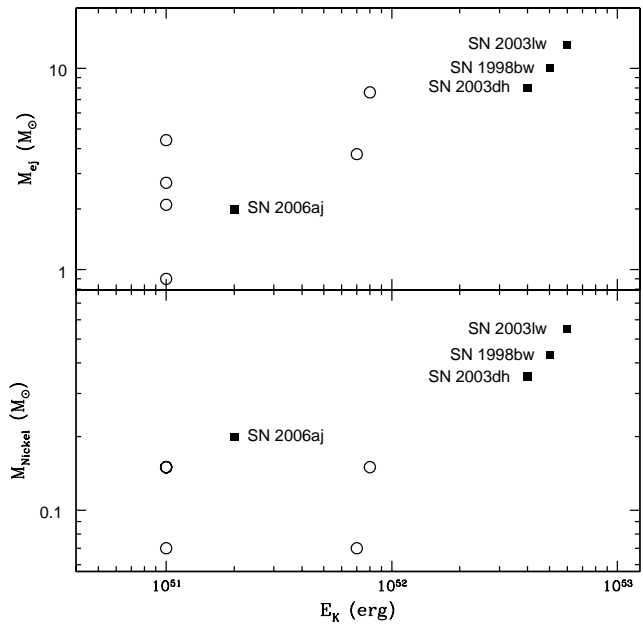}
\caption{The ejected mass ({\it upper panel}) and nickel yield ({\it lower 
panel}) of core-collapse SNe, versus the SN explosion energy. Filled squares 
are the four SNe with confirmed GRB-connection. Open circles are six Type Ibc 
SNe without an observational signature for association with GRBs, taken from 
\citet{ham04}. The two circles with explosion energy near $10^{52}$ erg 
are the ``hypernovae'' 1997ef (upper circle) and 2002ap (lower circle). 
The circle with the lowest ejected mass and nickel yield at $E_K = 10^{51}$ 
erg is the ``standard'' Type Ic SN 1994I. In the lower panel, the upper 
circle at $E_K = 10^{51}$ erg represents three SNe as they have identical 
explosion energy and nickel yield.  (For SNe 1998bw, 2003dh, and 2003lw, 
the value of $M_{\rm Nickel}$ is taken to be the mean of the upper and 
lower limits in Table~\ref{grbsn}.)
}
\label{ek_mej}
\end{figure}

Given its very soft spectrum and the under-energetic nature, and the fact that
the SN associated with it appears to have a moderate explosion energy and
ejected mass, we speculate that GRB 060218 is a marginal gamma-ray burst 
since it appears to be close to the bottom line of GRB-SN connection. This 
consideration is best illustrated in Fig.~\ref{ek_mej}, which shows the four 
GRB-connected SNe and six Type Ibc SNe with no detected GRB-connection in 
the ejected mass-explosion energy and the nickel mass-explosion energy plane. 
Clearly, SN 2006aj is closer to normal Type Ibc SNe than to the other three 
GRB-connected SNe.

SN 2003jd, discovered in MCG--01-59-021 at redshift $z=0.01886$, has been 
argued 
to be an evidence of aspherical explosion viewed from a direction near the 
equatorial plane, based on the observation of its double-peaked nebular lines 
of neutral oxygen and magnesium \citep{maz05}. This Type Ic supernova is only 
slightly less luminous than SN 1998bw but brighter than SN 2006aj, thus it 
has been anticipated that a GRB could have accompanied it but has not been
seen because of the off-axis nature. However, a radio observation on it
taken at $\sim 1.6$ yr after the explosion has detected no emission from an
off-axis jet, which has been used to argue against a GRB connection for
SN 2003jd \citep{sod06a}. Thus, SN 2003jd might be a violator of our 
equations~(\ref{ep_lp}) and (\ref{eiso_up}). However, our 
equation~(\ref{eiso_up}) only gives an upper bound on the isotropic energy
of the GRB. Another point is that \citet{sod06a} have only tested a single 
model for off-axis GRBs, their no-detection result may have just ruled out 
one specific model (P. Mazzali, private communication).

If equation~(\ref{ep_lp}) is extended to cosmological GRBs, a limit on the 
peak luminosity of the underlying SNe can be calculated. Applying to 
GRB 990123, which is at redshift $1.6$ and has the maximum determined 
intrinsic peak spectral energy $E_{\gamma,\p} \approx 2,000$ keV (and 
$E_{\gamma,\iso} = 2.66\times 10^{54}$ erg) \citep{ama02,ama06}, we get 
$L_{\sn,\p} \approx 1.87\times 10^{43} {\rm erg}\,{\rm s}^{-1}$, only two 
times brighter than SN 1998bw (while GRB 990123 is brighter than GRB 980425
by six orders of magnitude!).

If cosmological GRBs are also associated with SNe \citep{zeh04}, it would
be interesting to know
out to what a redshift would the SNe be detectable. This is a question that 
is not easy to answer because as the redshift increases the luminosity of the 
SN would be easily overshined by the afterglow of the GRB if the afterglow
is bright. For a SN that is as luminous as twice of SN 1998bw, the 
Ultra-Violet/Optical Telescope (UVOT) on board \sw\, would be able to detect 
it to a redshift of $\approx 0.7$ according to the sensitivity of UVOT 
$m_{\rm B} = 24.0$ in white light in 1,000~s \citep{rom05}. Since the 
luminosity of SN 1998bw is comparable to that of SNe Ia, it can be expected 
that the upcoming space observatory {\it SuperNova Acceleration Probe} 
({\it SNAP}) would be able to detect GRB-connected SNe to redshift $\sim 
1.7$ \citep{ald05} under favorable conditions (i.e., the afterglow of the 
GRB does not overshine the SN but the GRB is still detectable as in the 
lucky case of GRB 980425/SN 1998bw).

\section{Summary and Conclusions}
\label{concl}

We have found a strong correlation between the peak spectral energy of GRBs
and the peak bolometric magnitude (i.e., the peak luminosity) of their 
underlying SNe, based on the observational data of the four pairs of GRBs and 
SNe with spectroscopically confirmed connection (Fig.~\ref{mag_epeak}, 
eqs.~\ref{ep_mp} and \ref{ep_lp}). The Pearson linear correlation coefficient 
between $\log E_{\gamma,\p}$ (the peak spectral energy of GRBs) and 
$-M_{\sn,\p}$ (the peak bolometric magnitude of SNe) is $0.997$, corresponding 
to a probability $P = 0.003$ for zero correlation. Although the sample is 
limited by the small number of GRBs-SNe, we consider the result to be very 
suggestive because of the large correlation coefficient. 

Combined with the relation between the peak spectral energy and the 
isotropic equivalent energy of GRBs \citep{ama06}, the correlation that we
have found leads to a relation between the isotropic energy of a GRB and the 
peak bolometric luminosity of the underlying supernova (eq.~\ref{eiso_up}).
If a GRB is among the normal cosmological class (i.e., it has a normal total
gamma-ray energy), and is indeed associated with a supernova, then 
equation~(\ref{eiso_up}) would take the equal sign. If a GRB is sub-energetic, 
like some of the SN-connected GRBs, equation~(\ref{eiso_up}) gives an upper 
bound on the isotropic gamma-ray energy.

The slope of $\log E_{\gamma,\iso} - \log L_{\sn,\p}$ is extremely steep, 
which is $\approx 10$ by equation~(\ref{eiso_up}). This naturally describes 
the observational fact that GRBs have a very large diversity in properties 
(e.g., the isotropic equivalent energy) compared to SNe. 

Applying the relations that we have obtained (eqs.~\ref{ep_lp} and 
\ref{eiso_up}) to normal Type Ibc SNe which are not as luminous as SN 1998bw,
we found that the prompt emission from the potential GRBs associated with
them peaks in the soft X-ray and UV band, and the total gamma-ray energy of 
the bursts is extremely small. Hence, the bursts associated with normal SNe 
Ibc would be more appropriately qualified as soft X-ray transients, which
might be easier to detect with X-ray or UV detectors than with gamma-ray 
detectors.

Despite the fact that cosmological GRBs are typically more luminous than the
sub-energetic GRB 980425 by five orders of magnitude or more, the potential 
SNe associated with them are expected to be brighter than SN 1998bw only by
a factor $\sim 2$. Although it is hard to predict in a general case up to 
what a distance a SN
associated with a cosmological GRB can be observed, under favorable conditions
a GRB-connected SN that is as luminous as a SN Ia should be observable up to 
redshift $\sim 1.7$ with the upcoming {\it SNAP} space observatory.

Our results suggest that the critical parameter characterizing the GRB-SN 
connection is the large peak luminosity of the SNe, rather than the 
broad-lined spectra (or, equivalently, the large expansion velocity) and/or 
the huge explosion energy as commonly hypothesized
\citep[and references therein]{nom04,del06,woo06}. Given the general Ansatz 
that 
the SN luminosity at peak equals the power generated by the decay of \nk\, 
\citep{arn82,mae03,nom04,maz06,pia06}, our results may indicate that the mass 
of \nk\, produced in the SN explosion is a key physical factor for 
understanding the nature of the GRB-SN connection as well as the nature of 
GRBs 
(Fig.~\ref{n56_epeak}). Although a physical relation between the peak spectral
energy of GRBs and the mass of \nk\, of SNe cannot be established based only 
on the results in this paper, the following consideration may provide 
us a clue. Popular models of GRBs involve aspherical explosion of massive 
stars, where gamma-ray emission is produced along the axis of the explosion 
via a jet or a shock \citep{woo93,pac98a,mac01,woo06a}. Investigation on 
nucleosynthesis in aspherical SN explosions indicates that \nk\, is 
distributed also preferentially in the direction along the jet axis where 
the ejecta carry more kinetic energy and the shock is stronger \citep{mae02}. 

Finally, we remark that a SN has been claimed to be detected in the afterglow
of GRB 020903, an extremely soft burst at redshift $0.251$ 
\citep{sod05,bers06}. The peak spectral energy of GRB 020903 is $3.37\pm 
1.79$ keV. The isotropic gamma-ray energy is $(2.8\pm 0.7)\times 10^{49}$ 
erg, which makes GRB 020903 consistent with the $E_{\gamma,\p}-E_{\gamma,
\iso}$ relation \citep{ama06}. Equation~(\ref{ep_mp}) then predicts that the 
SN associated with GRB 020903 has a peak bolometric magnitude $\approx 
-18.06$, fainter than SN 1998bw by $\sim 0.6$ mag. Fitting the SN 1998bw
template to the bump in the afterglow lightcurve of GRB 020903, \citet{bers06} 
found that the SN is fainter than SN 1998bw by $0.8\pm 0.1$ mag at peak in 
the $R$-band, consistent with the $0.6\pm 0.5$ mag reported by \citet{sod05}
earlier. Considering the fact that the spectrum of the SN of GRB 020903 at 
38.6 days after the burst was redder than SN 1998bw, the difference in the
bolometric magnitudes is likely to be less than 0.8 mag (D. Bersier, private 
communication). It appears that GRB 020903 and its supernova are consistent 
with equations~(\ref{ep_lp}) and (\ref{eiso_up}), although a conclusion cannot 
be made without the availability of data in other filters.

\section*{Acknowledgments}

The author thanks D. Bersier, S. Campana, J. Deng, P. Mazzali, F. Patat, 
E. Pian, and S. Taubenberger for useful communications and sharing data. 
He also thanks the referee (P. Mazzali) for a wonderful report which has 
led to significant improvements to the paper.

\appendix

\section{The Peak Bolometric Magnitude of SN~2003dh}
\label{sn03dh}

\begin{figure}
\vspace{2pt}
\includegraphics[angle=0,scale=0.453]{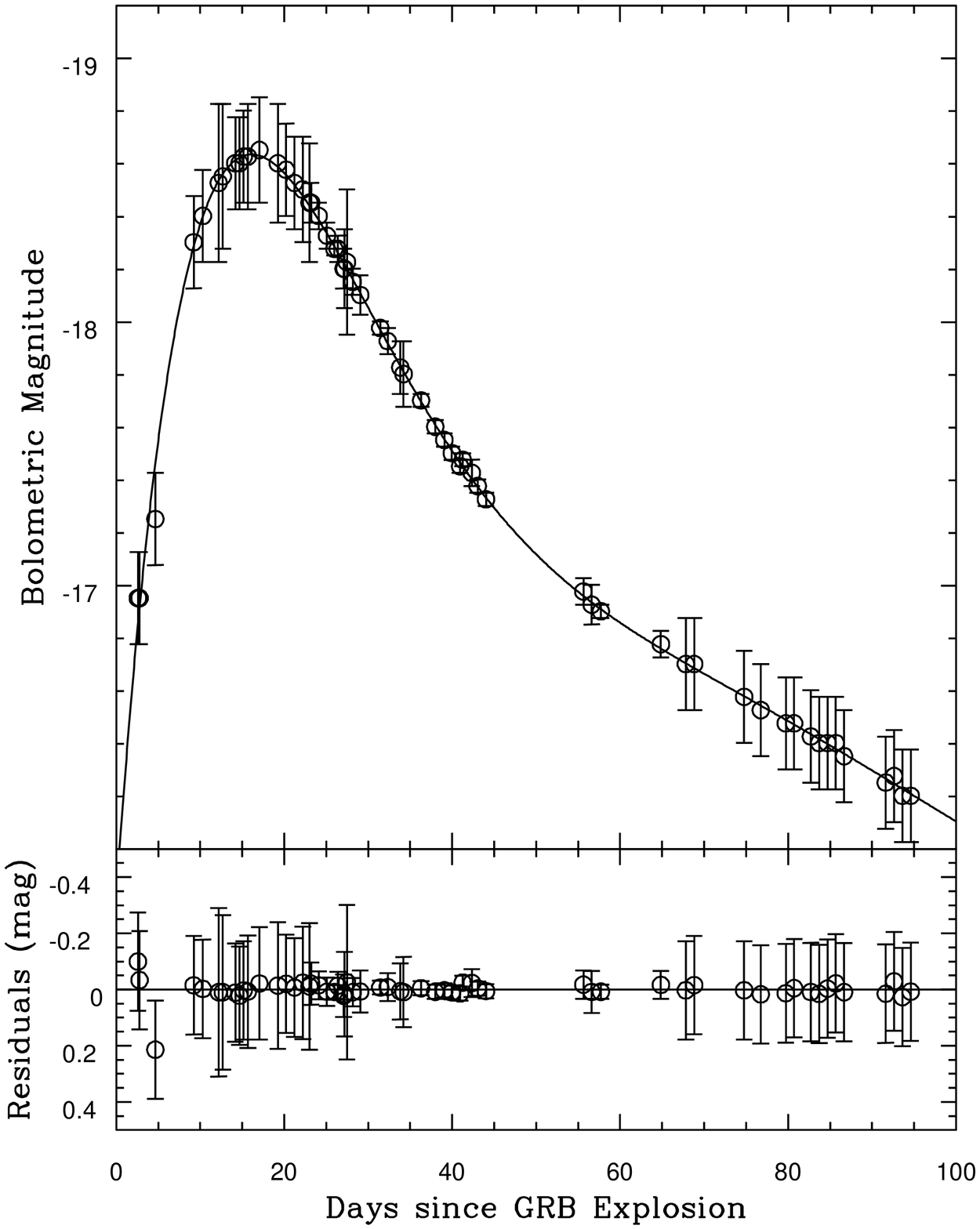}
\caption{{\it Upper panel:} Fitting the restframe bolometric lightcurve of 
SN 1998bw with a ninth-order polynomial. The best fit (the solid curve) has 
$\chi^2/\dof = 0.04$, with $\dof=86$. {\it Lower panel:} Residuals of the 
fit.
}
\label{template}
\end{figure}

The peak of the lightcurve of SN 2003dh was not captured \citep{den05}. To
obtain the peak magnitude of SN 2003dh, we fit the available lightcurve
data with a model-independent empirical approach.

The striking similarity between the spectrum of SN 2003dh and that of SN 1998bw
\citep{sta03} enables us to use the bolometric lightcurve of SN 1998bw,
which has been well sampled and studied \citep{pat01,nak01,maz06b,pia06}, as 
a template to fit the available light curve data of SN 2003dh. This has been 
a standard approach in searching for SNe in the optical afterglows of GRBs 
\citep{zeh04,sod05,bers06}.

First, we fit the restframe bolometric lightcurve of SN 1998bw with a 
polynomial. The data (the same as that used in Mazzali et al. 2006b and Pian 
et al. 2006) are kindly provided by E. Pian, which differ from that in
\citet{pat01} by a constant scaling factor in the bolometric luminosity. The 
bolometric luminosity of SN 1998bw in \citet{maz06b} is smaller than that in 
\citet{pat01} by a factor $\approx 0.83$ (i.e., $0.2$ mag fainter), resulted 
from the fact that a different cosmology and different reddening/extinction 
have been adopted in \citet{pat01}. In order to construct the 
template lightcurve, we use the lightcurve data from 2.5 day to 187 day after 
GRB 980425 in the restframe, which consist totally of 96 data points and span 
a time-interval that is large enough for 
the purpose here. We find that the lightcurve of SN 1998bw in the above time
range is best fitted by a ninth-order polynomial (Fig.~\ref{template}), with 
$\chi^2/\dof = 0.04$ 
($\dof = 86$).

Then, we take the smooth curve defined by the ninth order polynomial (the 
solid curve in Fig.~\ref{template}) as a template and fit it to the 
bolometric lightcurve of SN 2003dh. In doing so, we stretch the template 
lightcurve, and shift it in magnitude and time. That is, if we denote the 
template lightcurve in magnitude by $M_{\rm template}(t)$, we fit the 
lightcurve of SN 2003dh with a magnitude function
\begin{eqnarray}
	M(t) = M_{\rm template}(\alpha t+\beta) + M_0 \;, \label{mfit}
\end{eqnarray}
where $t$ is time, $\alpha$, $\beta$, and $M_0$ are parameters to be 
determined.

The bolometric lightcurve data of SN 2003dh are taken from \citet{den05}, 
rescaled to the cosmology adopted in \citet{maz06b} and this paper. In 
\citet{den05}, the luminosity distance of SN 2003dh was taken to be $809$ Mpc 
(i.e., distance modulus $= 39.54$). While in our cosmology, the luminosity 
distance of SN 2003dh is $791$ Mpc. Thus, the luminosity of SN 2003dh is 
reduced by a factor of $0.956$ (i.e., $0.05$ mag fainter), adopting the same 
reddening/extinction. 

\begin{figure}
\vspace{2pt}
\includegraphics[angle=0,scale=0.463]{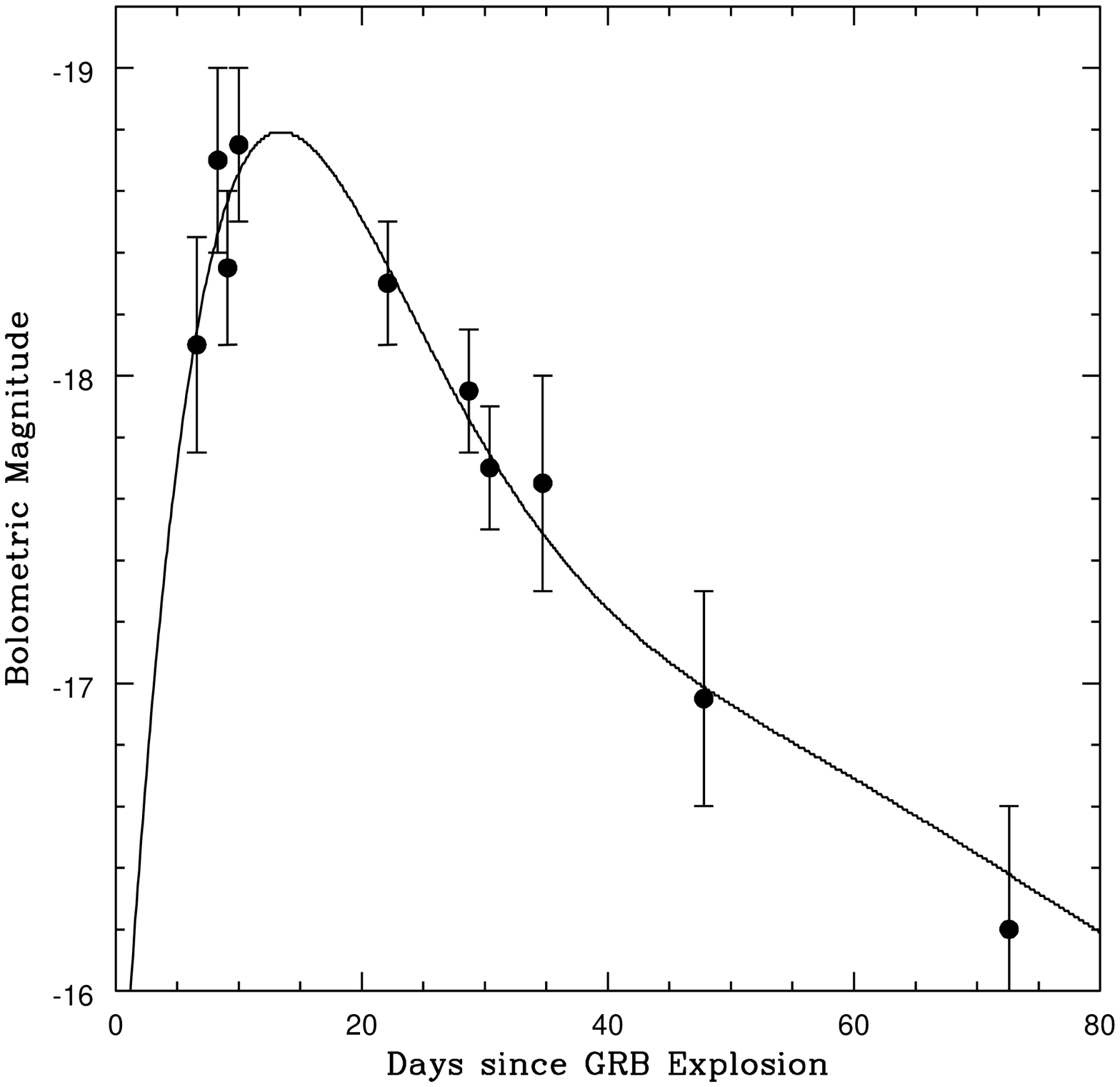}
\caption{Fitting the restframe bolometric lightcurve of SN 2003dh with a 
template of SN 1998bw defined by the solid curve in Fig.~\ref{template}, by 
stretching, rescaling the template lightcurve, and shifting its time origin 
(eq.~\ref{mfit}). The best fit (the solid curve) has $\chi^2/\dof = 0.3$, 
with $\dof = 7$. The peak magnitude given by the solid curve is $-18.79$, 
occurring at $13.4$ day after GRB 030329. 
}
\label{fit03dh}
\end{figure}

The results of fitting the template to the data of SN 2003dh are as follows: 
$\alpha = 1.32$, $\beta = -1.6$, and $M_0 = -0.16$ (Fig.~\ref{fit03dh}). The 
$\chi^2/\dof = 0.3$, where $\dof = 7$. We estimate the peak magnitude of 
SN 2003dh by the minimum of the $M(t)$ (the peak of the solid curve in 
Fig.~\ref{fit03dh}), which is $-18.79 \pm 0.23$. The peak occurs at $13.4$ 
day after GRB 030329 in the restframe, consistent with the $10-13$ day 
estimated by \citet{hjo03}.

Applying the procedure to SN 2003lw and SN 2006aj, we obtain results that
are consistent with the numbers listed in Table~\ref{grbsn}.

\bsp

\label{lastpage}


\begin{thebibliography}{99}

\bibitem[\protect\citeauthoryear{Aldering}{2005}]{ald05}
	Aldering G., 2005, New Astron. Rev., 49, 346

\bibitem[\protect\citeauthoryear{Amati}{2006}]{ama06}
	Amati L., 2006, astro-ph/0601553

\bibitem[\protect\citeauthoryear{Amati et al.}{2006}]{ama06b}
        Amati L., Della Valle M., Frontera F., Malesani D., Guidorzi C., 
	Montanari E., Pian E., 2006, astro-ph/0607148

\bibitem[\protect\citeauthoryear{Amati et al.}{2002}]{ama02}
        Amati L. et al., 2002, A\&A, 390, 81

\bibitem[\protect\citeauthoryear{Arnett}{1982}]{arn82}
	Arnett W. D., 1982, ApJ, 253, 785

\bibitem[\protect\citeauthoryear{Berger \& Becker}{2005}]{ber05}
	Berger E., Becker G., 2005, GCN 3520

\bibitem[\protect\citeauthoryear{Bersier et al.}{2006}]{bers06}
	Bersier D. et al., 2006, ApJ, 643, 284

\bibitem[\protect\citeauthoryear{Blinnikov et al.}{2002}]{bli02}
        Blinnikov S. I., Nadyozhin D. K., Woosley S. E., Sorokina
        E. I., 2002, in Nuclear Astrophysics, ed. W. Hillebrandt \&
        E. M\"uller (Garching: Max-Planck-Institut f\"ur Astrophysik),
        p. 144

\bibitem[\protect\citeauthoryear{Bloom, Frail \& Kulkarni}{2003}]{blo03}
	Bloom J. S., Frail D. A., Kulkarni S. R., 2003, ApJ, 594, 674

\bibitem[\protect\citeauthoryear{Bloom et al.}{1999}]{blo99}
	Bloom J. S. et al., 1999, Nature, 401, 453

\bibitem[\protect\citeauthoryear{Bloom et al.}{2002}]{blo02}
	Bloom J. S. et al., 2002, ApJ, 572, L45

\bibitem[\protect\citeauthoryear{Calzavara \& Matzner}{2004}]{cal04}
	Calzavara A. J., Matzner C. D., 2004, MNRAS, 351, 694

\bibitem[\protect\citeauthoryear{Campana et al.}{2006}]{cam06}
 	Campana S. et al., 2006, astro-ph/0603279

\bibitem[\protect\citeauthoryear{Cobb et al.}{2006}]{cob06}
	Cobb B. E., Bailyn C. D., van Dokkum P. G., Natarajan 
	P., 2006, ApJ, 645, L113

\bibitem[\protect\citeauthoryear{Colgate}{1968}]{col68}
	Colgate S. A., 1968, Canadian J. Phys., 46, S476

\bibitem[\protect\citeauthoryear{Della Valle}{2006}]{del06}
	Della Valle M., 2006, in Gamma-Ray Bursts in the Swift Era,
	Sixteenth Maryland Astrophysics Conference, ed. S. S. Holt, 
	N. Gehrels, \& J. A. Nousek (NY: American Institute 
	of Physics), p. 367

\bibitem[\protect\citeauthoryear{Deng et al.}{2005}]{den05}
	Deng J., Tominaga N., Mazzali P. A., Maeda K., Nomoto K., 2005,
	ApJ, 624, 898

\bibitem[\protect\citeauthoryear{Esin \& Blandford}{2000}]{esi00}
	Esin A. A., Blandford R., 2000, ApJ, 534, L151

\bibitem[\protect\citeauthoryear{Feldmeier, Ciardullo \& Jacoby}
	{1997}]{fel97}
	Feldmeier J. J., Ciardullo R., Jacoby G. H., 1997, ApJ, 
	479, 231

\bibitem[\protect\citeauthoryear{Ferrero et al.}{2006}]{fer06}
	Ferrero P. et al., 2006, astro-ph/0605058

\bibitem[\protect\citeauthoryear{Frail et al.}{2001}]{fra01}
	Frail D. A. et al., 2001, ApJ, 562, L55

\bibitem[\protect\citeauthoryear{Gal-Yam, Ofek \& Shemmer}{2002}]{gal02}
	Gal-Yam A., Ofek E. O., Shemmer O., 2002, MNRAS, 332, L73

\bibitem[\protect\citeauthoryear{Galama et al.}{1998}]{gal98}
        Galama T. J. et al., 1998, Nature, 395, 670

\bibitem[\protect\citeauthoryear{Garnavich et al.}{2003}]{gar03}
	Garnavich P. M. et al., 2003, ApJ, 582, 924

\bibitem[\protect\citeauthoryear{Ghirlanda, Ghisellini \& Lazzati}{2004}]{ghi04}
        Ghirlanda G., Ghisellini G., Lazzati D., 2004, ApJ, 616, 331

\bibitem[\protect\citeauthoryear{Greiner et al.}{2003}]{gre03}
	Greiner J. et al., 2003, ApJ, 599, 1223

\bibitem[\protect\citeauthoryear{Grupe et al.}{2006}]{gru06}
	Grupe D. et al., 2006, ApJ, 645, 464

\bibitem[\protect\citeauthoryear{Hamuy}{2004}]{ham04}
	Hamuy M., 2004, in Stellar collapse, ed. C. L. Fryer (Dordrecht: 
	Kluwer Academic Publishers), p. 39

\bibitem[\protect\citeauthoryear{Hjorth et al.}{2003}]{hjo03}
	Hjorth J. et al., 2003, Nature, 423, 847

\bibitem[\protect\citeauthoryear{Hurley et al.}{2002}]{hur02}
	Hurley K. et al., 2002, GCN 1252

\bibitem[\protect\citeauthoryear{Iwamoto et al.}{1998}]{iwa98}
	Iwamoto, K. et al., 1998, Nature, 395, 672

\bibitem[\protect\citeauthoryear{Iwamoto et al.}{2000}]{iwa00}
	Iwamoto, K. et al., 2000, ApJ, 534, 660

\bibitem[\protect\citeauthoryear{Lamb, Donaghy \& Graziani}{2005}]{lam05}
	Lamb D. Q., Donaghy T. Q., Graziani C., 2005, ApJ, 620, 355

\bibitem[\protect\citeauthoryear{Li}{2006}]{li06}
	Li L. -X., 2006, astro-ph/0605387

\bibitem[\protect\citeauthoryear{Liang, Zhang \& Dai}{2006}]{lia06}
	Liang E., Zhang B., Dai Z. G., 2006, astro-ph/0605200

\bibitem[\protect\citeauthoryear{MacFadyen, Woosley \& Heger}{2001}]{mac01}
	MacFadyen A. I., Woosley S. E., Heger A., 2001, ApJ, 550, 410

\bibitem[\protect\citeauthoryear{Maeda et al.}{2002}]{mae02}
	Maeda K., Nakamura T., Nomoto K., Mazzali P. A., Patat F., Hachisu I.,
	2002, ApJ, 565, 405

\bibitem[\protect\citeauthoryear{Maeda et al.}{2003}]{mae03}
	Maeda K., Mazzali P. A., Deng J., Nomoto K., Yoshii Y., Tomita 
	H., Kobayashi Y., 2003, ApJ, 593,931

\bibitem[\protect\citeauthoryear{Malesani et al.}{2004}]{mal04}
	Malesani D. et al., 2003, ApJ, 609, L5

\bibitem[\protect\citeauthoryear{Masetti et al.}{2006}]{mas06}
	Masetti N., Palazzi E., Pian E., Patat F., 2006, GCN 4803

\bibitem[\protect\citeauthoryear{Mazzali et al.}{2000}]{maz00}
	Mazzali P. A., Iwamoto K., Nomoto K., 2000, ApJ, 545, 407

\bibitem[\protect\citeauthoryear{Mazzali et al.}{2004}]{maz04}
	Mazzali P. A., Deng J., Maeda K., Nomoto K., Filippenko A. V., 
	Matheson T., 2004, ApJ, 614, 858;

\bibitem[\protect\citeauthoryear{Mazzali et al.}{2006a}]{maz06}
	Mazzali P. A., Deng J., Nomoto K., Pian E., Tominaga N., 
	Tanaka M., Maeda K., 2006a, astro-ph/0603567

\bibitem[\protect\citeauthoryear{Mazzali et al.}{2002}]{maz02}
	Mazzali P. A. et al., 2002, ApJ, 572, L61

\bibitem[\protect\citeauthoryear{Mazzali et al.}{2005}]{maz05}
	Mazzali P. A. et al., 2005, Science, 308, 1284

\bibitem[\protect\citeauthoryear{Mazzali et al.}{2006b}]{maz06b}
	Mazzali P. A. et al., 2006b, ApJ, 645, 1323

\bibitem[\protect\citeauthoryear{Mirabal et al.}{2006}]{mir06}
	Mirabal N., Halpern J. P., An D., Thorstensen J. R., Terndrup
	D. M., 2006, ApJ, 643, L99

\bibitem[\protect\citeauthoryear{Modjaz et al.}{2006}]{mod06}
	Modjaz M. et al., 2006, ApJ, 645, L21

\bibitem[\protect\citeauthoryear{Nakamura et al.}{2001}]{nak01}
	Nakamura T., Mazzali P. A., Nomoto K., Iwamoto K., 2001,
	ApJ, 550, 991

\bibitem[\protect\citeauthoryear{Nakar \& Piran}{2005}]{nak05}
	Nakar E., Piran T., 2005, MNRAS, 360, L73

\bibitem[\protect\citeauthoryear{Nomoto et al.}{2004}]{nom04}
	Nomoto K., Maeda K, Mazzali P. A., Umeda H., Deng J., Iwamoto K., 
	2004, in Stellar collapse, ed. C. L. Fryer (Dordrecht: 
	Kluwer Academic Publishers), p. 277

\bibitem[\protect\citeauthoryear{Paczy\'nski}{1998a}]{pac98a}
	Paczy\'nski B., 1998a, ApJ, 494, L45

\bibitem[\protect\citeauthoryear{Paczy\'nski}{1998b}]{pac98b}
	Paczy\'nski B., 1998b, in Gamma-Ray Bursts: 4th Huntsville 
	Symposium, ed. C. A. Meegan, R. D. Preece, \& T. M. Koshut
	(New York: American Institute of Physics), p. 783

\bibitem[\protect\citeauthoryear{Patat et al.}{2001}]{pat01}
	Patat F. et al., 2001, ApJ, 555, 900

\bibitem[\protect\citeauthoryear{Pian et al.}{2006}]{pia06}
	Pian E. et al., 2006, astro-ph/0603530

\bibitem[\protect\citeauthoryear{Piran}{2004}]{pir04}
	Piran T., 2004, Rev. Mod. Phys., 76, 1143

\bibitem[\protect\citeauthoryear{Podsiadlowski et al.}{2004}]{pod04}
	Podsiadlowski Ph., Mazzali P. A., Nomoto K., Lazzati D., 
	Cappellaro E., 2004, ApJ, 607, L17

\bibitem[\protect\citeauthoryear{Preece et al.}{2000}]{pre00}
	Preece R. D., Briggs M. S., Mallozzi R. S., Pendleton G. N., 
	Paciesas W. S., Band, D. L., 2000, ApJS, 126, 19

\bibitem[\protect\citeauthoryear{Prochaska et al.}{2004}]{pro04}
	Prochaska J. X. et al., 2004, ApJ, 611, 200

\bibitem[\protect\citeauthoryear{Racusin et al.}{2005}]{rac05}
	Racusin J. et al., 2005, GCN 4169

\bibitem[\protect\citeauthoryear{Ramirez-Ruiz et al.}{2005}]{ram05}
	Ramirez-Ruiz E., Granot J., Kouveliotou C., Woosley S. E., 
	Patel S. K., Mazzali P. A., 2005, ApJ, 625, L91

\bibitem[\protect\citeauthoryear{Roming et al.}{2005}]{rom05}
	Roming P. W. A. et al., 2005, Space Sci. Rev., 120, 95 

\bibitem[\protect\citeauthoryear{Sakamoto et al.}{2005}]{sak05}
	Sakamoto T. et al., 2005, ApJ, 629, 311

\bibitem[\protect\citeauthoryear{Sari, Piran \& Halpern}{1999}]{sar99}
	Sari R., Piran T., Halpern J. P., 1999, ApJ, 519, L17

\bibitem[\protect\citeauthoryear{Sauer et al.}{2006}]{sau06}
	Sauer D. N., Mazzali P. A., Deng J., Valenti S., Nomoto K.,
	Filippenko A. V., 2006, MNRAS, 369, 1939

\bibitem[\protect\citeauthoryear{Sazonov, Lutovinov \& Sunyaev}{2004}]{saz04}
	Sazonov S. Yu., Lutovinov A. A., Sunyaev R. A., 2004, Nature,
	430, 646

\bibitem[\protect\citeauthoryear{Soderberg, Frail \& Wieringa}{2004}]{sod04a}
	Soderberg A. M., Frail D. A., Wieringa M. H., 2004, ApJ, 607, L13

\bibitem[\protect\citeauthoryear{Soderberg et al.}{2006a}]{sod06a}
	Soderberg A. M., Nakar E., Berger E., Kulkarni S. R. 2006a, ApJ,
	638, 930

\bibitem[\protect\citeauthoryear{Soderberg et al.}{2004}]{sod04}
	Soderberg A. M. et al., 2004, Nature, 430, 648

\bibitem[\protect\citeauthoryear{Soderberg et al.}{2005}]{sod05}
	Soderberg A. M. et al., 2005, ApJ, 627, 877

\bibitem[\protect\citeauthoryear{Soderberg et al.}{2006b}]{sod06}
	Soderberg A. M. et al., 2006b, astro-ph/0604389

\bibitem[\protect\citeauthoryear{Sollerman et al.}{2006}]{sol06}
	Sollerman J. et al., 2006, A\&A, 454, 503

\bibitem[\protect\citeauthoryear{Stanek et al.}{2003}]{sta03}
        Stanek et al., 2003, ApJ, 591, L17

\bibitem[\protect\citeauthoryear{Stanek et al.}{2005}]{sta05}
	Stanek et al., 2005, ApJ, 626, L5

\bibitem[\protect\citeauthoryear{Taubenberger et al.}{2006}]{tau06}
	Taubenberger S. et al., 2006, astro-ph/0607078

\bibitem[\protect\citeauthoryear{Tomita et al.}{2006}]{tom06}
	Tomita H. et al., 2006, ApJ, 644, 400

\bibitem[\protect\citeauthoryear{Ulanov et al.}{2005}]{ula05}
	Ulanov M. V., Golenetskii S. V., Frederiks D. D., Aptekar R. L,
	Mazets E. P., Kokomov A. A., Palshin V. D., 2005, Nuovo Cimento 
	C., 28, 351

\bibitem[\protect\citeauthoryear{van der Horst et al.}{2005}]{hor05}
	van der Horst A. J., Rol E., Wijers R. A. M. J., Strom R., Kaper 
	L., Kouveliotou C., 2005, ApJ, 634, 1166

\bibitem[\protect\citeauthoryear{Vaughan et al.}{2006}]{vau06}
	Vaughan S. et al., 2006, ApJ, 638, 920

\bibitem[\protect\citeauthoryear{Wang \& Wheeler}{1998}]{wan98}
	Wang L., Wheeler J. C., 1998, ApJ, 504, L87

\bibitem[\protect\citeauthoryear{Waxman}{2004}]{wax04}
	Waxman E., 2004, ApJ, 602, 886

\bibitem[\protect\citeauthoryear{Waxman \& Draine}{2000}]{wax00}
	Waxman E., Draine B. T., 2000, ApJ, 537, 796

\bibitem[\protect\citeauthoryear{Woosley}{1993}]{woo93}
        Woosley S. E., 1993, ApJ, 405, 273

\bibitem[\protect\citeauthoryear{Woosley \& Heger}{2006a}]{woo06a}
        Woosley S. E., Heger A., 2006a, ApJ, 637, 914

\bibitem[\protect\citeauthoryear{Woosley \& Heger}{2006b}]{woo06}
	Woosley	S. E., Heger A., 2006b, in Gamma-Ray Bursts in the Swift Era,
	Sixteenth Maryland Astrophysics Conference, ed. S. S. Holt, 
	N. Gehrels, \& J. A. Nousek (NY: American Institute 
	of Physics), p. 398

\bibitem[\protect\citeauthoryear{Yamazaki, Yonetoku \& Nakamura}{2003}]{yam03}
	Yamazaki R., Yonetoku D., Nakamura T., 2003, ApJ, 594, L79

\bibitem[\protect\citeauthoryear{Zeh, Klose \& Hartmann}{2004}]{zeh04}
	Zeh A., Klose S., Hartmann D. H., 2004, ApJ, 609, 952

\end{thebibliography}
\end{document}